# Radiomics in Cancer Radiotherapy: A Review

Jiwoong Jeong[1], Arif Ali[1], Tian Liu[1], Hui Mao[2], Walter J. Curran[1], Xiaofeng Yang[1*]

[1] *Department of Radiation Oncology and Winship Cancer Institute, Emory University, Atlanta, GA 30322*

[2] *Department of Radiology and Imaging Sciences and Winship Cancer Institute, Emory University, Atlanta, GA 30322*

*Corresonding to: xiaofeng.yang@emory.edu

## Abstract

Radiomics is a nascent field in quantitative imaging that uses advanced algorithms and considerable computing power to describe tumor phenotypes, monitor treatment response, and assess normal tissue toxicity quantifiably. Remarkable interest has been drawn to the field due to its noninvasive nature and potential for diagnosing and predicting patient prognosis. This review will attempt to comprehensively and critically discuss the various aspects of radiomics including its workflow, applications to different modalities, potential applications in cancer radiotherapy, and limitations.

## 1. Introduction

Cancer can be diagnosed and monitored through many different measures such as lab tests, biopsies, and imaging procedures. Minimally invasive lab tests specifically tailored for each cancer can monitor cancer biomarkers such as prostate-specific antigens (PSAs) (1), alpha-fetoprotein (AFP), or carcinoembryonic antigen (CEA) in the blood that might signify cancer progression or treatment response. The dilemma of observing biomarker levels only through laboratory testing is that it is just one measure of specific cancer cells that should be supplemented by other sources of information such as biopsies and medical images.

The standard method to identify and characterize tumors based on genetic and pathologic factors is through tissue biopsies. Biopsies and associated genetic and histopathologic analyses are able to diagnose and provide more information about the cancer but also have limitations. Most tumors are spatially and temporally heterogeneous due to irregularities in metabolism, vasculature, oxygenation, and gene expression. Such heterogeneities can require multiple biopsies to acquire a complete assessment of the tumor, which increases the risk of complications for the patient. However, the use of advanced medical imaging to supplement traditional diagnostic methods has the potential to better assess spatial and temporal tumor dynamics.

Medical imaging including computed tomography (CT), magnetic resonance imaging (MRI), positron emission tomography (PET), and ultrasound (US) can sample the entire tumor non-invasively multiple times while retaining the cellular and genetic information derived from phenotypic patterns. Although medical imaging is an ever advancing and powerful tool, many visible features derived from images are still being assessed qualitatively by humans. These assessments are often quite variable and subjective (2, 3). However, that is not to say that quantitative imaging (QI) methods such as dynamic susceptibility contrast enhanced (DSC) MRI (4) and MRI spectroscopy (5), that quantifies the amount of contrast or biochemical properties within the field of view (FOV) to evaluate blood perfusion and metabolism are not available. QI methods have the potential to provide a greater insight in the underlying tumor biology and could directly facilitate personalized treatments but are in the process of being integrated into clinical radiation oncology practice and will not be the focus of this review (6). Advancements in medical imaging is currently challenged by the limits of human observation and may be improved with the introduction of more objective and robust methods of extracting features from medical images.

Radiomics involves the high-throughput extraction of advanced quantitative features invisible to the naked eye to objectively and quantitatively describe tumor phenotypes in medical imaging. These radiomics features are 'mined' from medical images using advanced algorithms that examine multiple dimensions of the image such as physical (shape and size), textural (the spatial arrangement of voxels) features, histogram-based, and filtered-based. The papers included in this review were derived from an initial search of the key word "radiomics" in PubMed in the last ten years. Once a considerable amount of papers were gathered, the relevant sources from these papers were also added to a manual review for a more complete view of the field of radiomics. Since other reviews of radiomics have shown the detailed workflow and much greater data extraction from medical images paving way for more personalized cancer therapy, this review will compliment others by providing a big picture look at each major imaging modality and cancer sites as well as discuss the future potential and limitations of radiomics.

## 2. Workflow of Radiomics

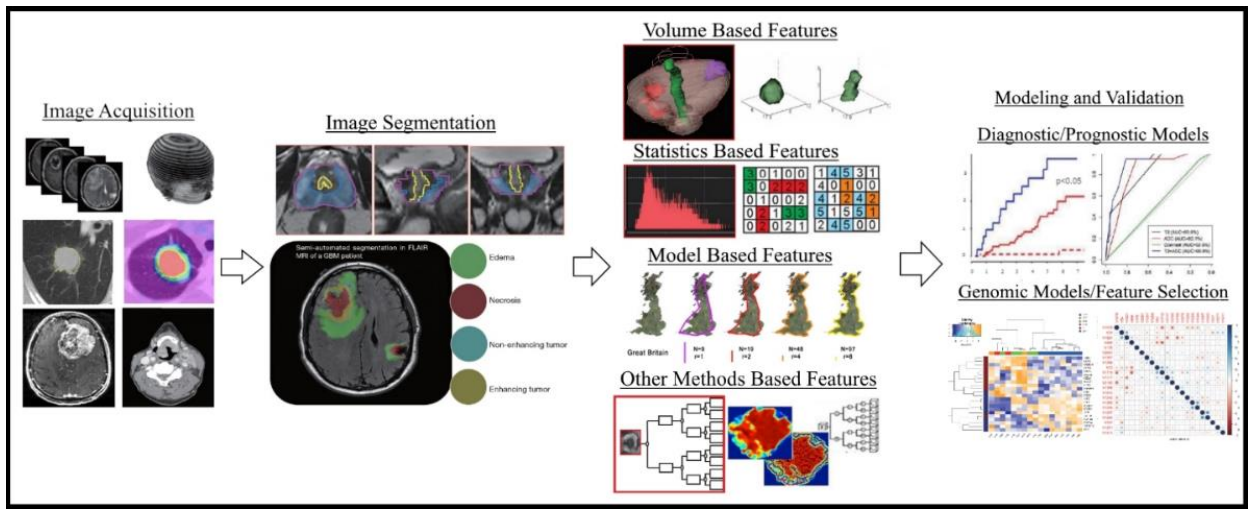

**Figure 1.** The workflow of Radiomics. The key components of the radiomic workflow is shown: Image acquisition and segmentation, feature extraction and analysis, and model testing and validation.

Current treatment decisions are based on a wide variety of traditional diagnostic tests. In recent years, radiomics has gained ground as a method in which one can predict and associate clinical outcomes of many patients with cancers like glioblastoma multiforme (GBM) (7), breast cancer (8), renal cell cancer (9), and head and neck cancer (10). Any medical image can be used to derive radiological features and associate them with clinical value like disease diagnosis, progression, and overall survival. Simply put, radiomics is an advanced computational identification, diagnosis, and prediction of patient response through medical images. As such, it has three key components: 1) image, 2) analysis, and 3) validation. These components are fundamental to the workflow of radiomics (Figure 1) and are complex processes that involve many different parts.

**2.1 Imaging**

Non-invasive medical imaging, like CT, MRI, PET, US, allows healthcare professionals to interrogate the image volume for any irregularities like tumors. Each modality has its advantages such as PET's ability to track glucose metabolism through radio-tracer uptake, MRI's ability to discriminate various tissue and image tissue function through various pulse sequences, and CT's asset in having electron density information for radiation therapy and dose calculations. To exploit these advantages and extract the most amount of relevant information, appropriate images are required for radiomic analysis.

Broadly speaking, the imaging component of radiomics can be separated into two parts: 1) image acquisition and 2) image segmentation (11). Images are acquired through various physical processes, detection parameters, and reconstruction algorithms to create a two or three dimensional patient image. Then the image is segmented into the regions or volumes of interest to focus on the cancerous areas as well as cut down on computational workload.

## 2.2 Analysis

Once the image acquisition and segmentation is applied to an initial patient data set, also called the training set, it is then analyzed through different algorithms for features of statistical significance. The goal of radiomic analysis is to find unique radiomic features or groups of features, called a signature, for various sites, malignancies, and outcomes to allow a semi-automatic or automatic approach to more informed clinical decisions. These analytical methods often describe the image in a way humans can't and allows algorithms to "mine" for data that may be correlated with clinical value. Methods can vary from histogram-based first order, co-occurrence-based second order, to matrix-based higher order texture analysis for MRI (12) and ultrasound (13). Texture analyses have been shown to predict chemotherapy response in non-small cell lung cancer (NSCLC) (14), differentiate prostate cancer by Gleason score (15), and show prognostic power in GBM (16). These features and signatures can be applied to even more detailed clinical applications such as tumor heterogeneity in glioblastomas (17) and tumor phenotype information (18, 19).

## 2.3 Validation

Once a feature or signature has been extracted from an image, it must be validated with another cohort, called a validation set, which is distinct from the original training set. The constant testing and retesting allows the researcher to strip away the redundant, poorly performing, or unstable features to get the best features that correlate with specific clinical parameters such as patient prognosis or tumor response to treatment as demonstrated in oropharyngeal, colorectal, and lung cancers (19-22).

**3. Review of Radiomics by Modality**

This paper summarizes various features into 4 categories as shown in Table 1: volume-based descriptors (C1), statistics based descriptors (C2), model based descriptors (C3), and other descriptors (C4). Volume based descriptors are features regarding the physical tumor size, shape, location, etc. Statistics based descriptors are varying orders of statistical outputs that describe the image such as $1^{st}$ order (C2a) statistics (comparing the pixel individually) like mean, standard deviation, kurtosis, skewness, etc.; $2^{nd}$ order (C2b) statistics (comparing pairs of pixels) like contrast, entropy, inertia, homogeneity, etc.; and higher order (C2c) statistics (comparing multiple pixels) like grey-level run length matrices. Model based descriptors are features derived from models, like fractal models, that describe an image (23) consistent with tumor progression. Finally, other method based descriptors are features derived from some processing of the raw or reconstructed imaging data like Nagakami parameters in ultrasound (24, 25) or Gabor filters (26).

Table 1. Radiomics feature categories

| Category | Name | Descriptors and examples |
|---|---|---|
| C1 | Volume-based | Tumor size, shape, location, etc. |
| C2 | Statistics-based | |
| C2a | 1st order statistics | Mean, median, standard deviation, kurtosis, skewness, quartiles, min/max, etc. |
| C2b | 2nd order statistics | Contrast, energy, entropy, correlation, inertia, cluster prominence, cluster shade, homogeneity, dissimilarity, etc. |
| C2c | Higher order statistics | Grey-level run length matrices, grey-level size zone matrices, neighborhood gray-tone difference matrices, etc. |
| C3 | Model-based | Fractal analysis |
| C4 | Other methods | Wavelets, Gabor transform, Nakagami parameters, contourlet, etc. |
| C5 | Signatures | Four or more features used together or separately |

Radiomics has become a hot topic in recent years. As seen in Fig. 1, over 78% of the papers with the radiomic features outlined in Table 1, were published in the last 5 years, from 2012 to 2016. This may be a response to the increased interest in personalized medicine, enhanced computing power, and large databases to extract information from. A list of papers containing the features in Table 1 is sorted by modality in Table 2 and their distribution by modality is summarized in Figure 2a and 2b. When assessing these papers as a whole, the big three modalities of CT, MRI, and PET account for 74% of the papers published. Additionally, 56% are focused on three specific sites: brain, head and neck, and lung which are strongly associated with MRI, PET, and CT respectively. Overall, statistics based features usually referred to as texture features, C2a, C2b, and C2c, by far are the most common while the use of signatures have become more common with about 73% of papers using signatures being published in the last 2 years, 2015 and 2016.

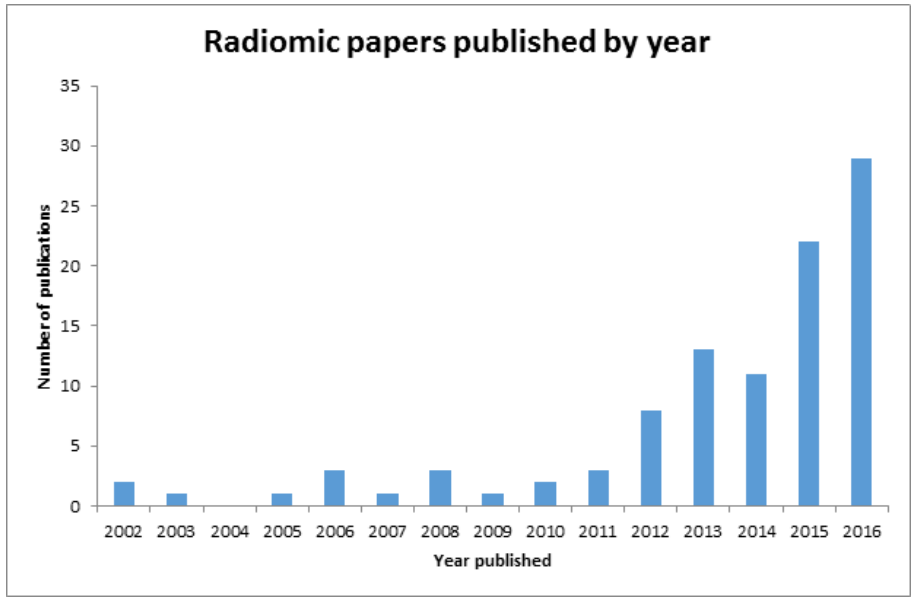

**Figure 2.** The number of radiomic study publications by year of publication.

**Table 2.** List of Radiomic Feature Categories by Modality.

| Modality | Author | Year | General Site | Specific Cancer/Disease | Feature Category |
|---|---|---|---|---|---|
| 4DCT | Fave | 2017 | Lung | NSCLC | Signature |
| CBCT | Fave | 2015 | Lung | NSCLC | C2 |
| CE-CT | Li | 2014 | Liver | Hepatocellular carcinoma | C1 and C4 |
| CE-CT | Petkovska | 2006 | Lung |  | C2a |
| CT | Aerts | 2014 | Head and Neck /Lung |  | Signature |
| CT | Aerts | 2016 | Lung | NSCLC | C2b |
| CT | Balagurunathan | 2014 | Lung | NSCLC | C1 |
| CT | Balagurunathan | 2014 | Lung | NSCLC | C2c |
| CT | Coroller | 2015 | Lung | NSCLC | Signature |
| CT | Cunliffe | 2015 | Lung | Radiation pneumonitis | Signature |
| CT | Fave | 2015 | Lung | NSCLC | C2a and C2b |
| CT | Ganeshan | 2010 | Lung | NSCLC | C2 |
| CT | Goh | 2011 | Kidneys | Renal Cell | C2 |
| CT | Hayano | 2016 | Lung | NSCLC | C1 and C2b |
| CT | Huynh | 2016 | Lung | NSCLC | C1, C2c, and C4 |
| CT | Huynh | 2016 | Lung | NSCLC | Signature |
| CT | Kido | 2002 | Lung |  | C3 |

| Modality | Author | Year | Site | Type | Category |
|---|---|---|---|---|---|
| CT | Kim | 2005 | Liver | Hepatocellular carcinoma | C2b |
| CT | Leijenaar | 2015 | Head and Neck | Oropharyngeal | Signature |
| CT | Liang | 2016 | Colorectal | | Signature |
| CT | McNitt-Gray | 1999 | Lung | | C2b |
| CT | Parmar | 2015 | Head and Neck /Lung | | Signature |
| CT | Permuth | 2016 | Pancreas | Pancreatic intraductal papillary mucinous neoplasms | Signature |
| CT | Rao | 2016 | Colorectal | Liver Metastasis | C2 |
| CT | Tateishi | 2002 | Lung | | C2a |
| CT | Tian | 2015 | Soft Tissue Sarcoma | | C2 |
| CT | Way | 2006 | Lung | | Signature |
| CT | Yamamoto | 2014 | Lung | NSCLC | Signature |
| CT | Yip | 2015 | Head and Neck | Esophageal | C2a and C2b |
| CT (DECT) | Choi | 2016 | Lung | NSCLC | C1, C2a, and C2b |
| CT/MRI | Prasanna | 2016 | Multiple | | C2c |
| DCE-MRI | Johansen | 2009 | Breast | | C2a |
| DCE-MRI | Peng | 2013 | Brain | | C2b |
| DCE-MRI | Shukla | 2012 | Head and Neck | | C2b |
| DW-MRI | King | 2013 | Head and Neck | Squamous cell carcinoma | C2a |
| Light microscopy | Colen | 2016 | Brain | CNS | Signature |
| MRI | Baek | 2012 | Brain | Glioblastoma multiforme | C2a |
| MRI | Cameron | 2016 | Prostate | | Signature |
| MRI | Coroller | 2016 | Brain | Meningioma | Signature |
| MRI | Diehn | 2008 | Brain | Glioblastoma multiforme | C1 |
| MRI | Ellingson | 2013 | Brain | Glioblastoma multiforme | C1 |
| MRI | Ginsburg | 2016 | Prostate | | C2 and C4 |
| MRI | Gnep | 2017 | Prostate | | C2b |
| MRI | Gutman | 2015 | Brain | Glioblastoma multiforme | C1 |
| MRI | Khalvati | 2015 | Prostate | | Signature |
| MRI | Kickingerede | 2016 | Brain | Glioblastoma multiforme | Signature |
| MRI | Kjaer | 1995 | Brain | | C2 |
| MRI | Lerski | 1993 | Brain | | C2 |
| MRI | Li | 2016 | Breast | | C1 and C2b |
| MRI | Lopez | 2016 | Brain | Glioblastoma | C1 |
| | | | | multiforme | |
| MRI | Mahmoud-Ghoneim | 2003 | Brain | | C2 |
| MRI | Naeini | 2013 | Brain | Glioblastoma multiforme | C1 |
| MRI | Nie | 2008 | Breast | | C1, C2a, and C2b |
| MRI | Nie | 2016 | Rectum | Rectal Cancer | C2b |
| MRI | Obeid | 2016 | Breast | Adipokines (fat specific cytokines) | C2a |
| MRI | Prasanna | 2016 | Brain | Glioblastoma multiforme | C2b |
| MRI | Rios Velazquez | 2016 | Breast | | C1 and C2b |
| MRI | van den Burg | 2016 | Ear | Meniere's Disease | Signature |
| MRI | Wang | 2015 | Breast | Parenchyma of the breast | C2a |
| MRI | Grossmann | 2016 | Brain | Glioblastoma multiforme | C2c |
| MRI (Diffusion) | Foroutan | 2013 | Bone | | C2a |
| PET | Brooks | 2014 | Cervical | | C1 |
| PET | Cheng | 2015 | Head and Neck | Oropharyngeal | C4 |
| PET | Cook | 2013 | Lung | NSCLC | C2b |
| PET | Dong | 2013 | Head and Neck | Esophageal | C2a and C2b |
| PET | Doumou | 2015 | Head and Neck | Esophageal | C2b |
| PET | Eary | 2008 | Sarcoma | | C2b |
| PET | Galavis | 2010 | Multiple | Adrenal gland carcinoma, lung, epiglottis, and esophagus cancer | C2b and C2c |
| PET | Grootjans | 2016 | Lung | | C2c |
| PET | Haberkorn | 1994 | Animal Study* | Spontaneous Mammary fibroadenoma, chemically-induced mammary adenocarcinoma and dunning prostate adenocarcinoma | C2a |
| PET | Hatt | 2013 | Head and Neck | Esophageal | C2b and C2c |
| PET | Hatt | 2015 | Multiple | | C1 and C2b |
| PET | Henriksson | 2007 | Animal Study* | Head and neck squamous cell carcinoma | C2a |
| PET | Higashi | 1993 | Adenocarcinoma | | C2a |
| PET | Higgins | 2012 | Head and Neck | | C2a |
| PET | Leijenaar | 2013 | Lung | NSCLC | Signature |
| PET | Mu | 2015 | Cervical | | C2c |

| Modality | Author | Year | Site | Cancer Type | Category |
|---|---|---|---|---|---|
| PET | Nair | 2012 | Lung | NSCLC | C2a |
| PET | Nyflot | 2015 | Phantom* | | C2c |
| PET | Orlhac | 2014 | Multiple | Metastatic colorectal cancer, non-small cell lung cancer, and breast cancer. | C2a |
| PET | Rahmim | 2016 | Multi | Oropharyngeal and pancreatic | C2 |
| PET | Rizk | 2006 | Head and Neck | Esophageal | C2a |
| PET | Tixier | 2011 | Head and Neck | Esophageal | C2b |
| PET | Tixier | 2012 | Head and Neck | Esophageal | C2b |
| PET | Tixier | 2014 | Lung | NSCLC | C2 |
| PET | Yang | 2015 | Cervical | | C2b and C2c |
| PET | Yip | 2016 | Head and Neck | Esophageal | C2b and C2c |
| PET/CT | Coroller | 2016 | Lung | NSCLC | Signature |
| PET/CT | Cortes-Rodicio | 2016 | Liver | | C2c |
| PET/CT | Hatt | 2011 | Lung | NSCLC | C1 and C2b |
| PET/MRI | Antunes | 2016 | Kidneys | Renal Cell | C2a |
| PET/MRI | Vallieres | 2015 | Lung | Soft tissue sarcomas | Signature |
| PET /PET-CT | Didierlaurent | 2012 | *N/A | | C2a |
| PET /PET-CT | Tan | 2013 | Head and Neck | Esophageal | C2a and C2b |
| PET-CT | Bundschuh | 2014 | Rectum | Rectal Cancer | C2 |
| PET-CT | Cheng | 2013 | Head and Neck | Oropharyngeal | C2b |
| PET-CT | Xu | 2014 | Bone and soft tissue | Malignant and benign | C2b |
| PET-CT | Yang | 2013 | Cervical | | C2 |
| PET-CT /4DPET | Huang | 2013 | Head and Neck /Lung | | C2a |
| PET-CT /DCE-CT | Tixier | 2014 | Colorectal | | C2b |
| SPECT/CT | Bowen | 2016 | Liver | Hepatocellular carcinoma | C1 |
| US | Yang | 2014 | Head and Neck | | C4 |
| US | Yang | 2015 | Head and Neck | | C4 |
| US | Yang | 2012 | Head and Neck | | C2a |
| US | Yang | 2012 | Head and Neck | | C2b |
| US | Zhang | 2015 | Breast | | C4 |
| US | Zhang | 2015 | Breast | | C4 |
| US | Zhang | 2017 | Breast | | C4 |

*non-human studies

## 3.1 Computed Tomography

Computed tomography (CT) scan is the most common medical imaging modality in cancer therapy. It is used as a diagnostic tool for tumor assessment, treatment planning, dose calculations, as well as a quality assurance of patient position before treatment. The near-universal use of CT scans for patients undergoing cancer treatments is a tremendous asset for the field of radiomics. The wealth of CT data available for analysis as well as the ease of application makes radiomic analysis of CT images desirable. A review of the literature shows that out of all cancer sites, lung cancer utilizes CT imaging the most.

One of the earliest papers published for the use of radiomics in lung cancer was in 1999. McNitt-Gray et. al. *(27)*, showed that two different statistical features, correlation and difference entropy, were able to differentiate between benign and malignant solitary pulmonary nodules consistently. Other studies have supported the ability to use radiomic features to differentiate pulmonary nodules through different contrast enhancement features (28) and fractal analysis (29, 30). The Kido *et. al* paper suggested that through binary and gray scale fractal dimension calculations it may be possible to differentiate between benign and malignant pulmonary nodules as well as adenocarcinomas from squamous cell carcinomas. Fractal analysis has also been used in a hepatocellular carcinoma study where patients with favorable progression-free survival (PFS) showed lower fractal dimension on anterior-posterior (AP) contrast enhanced CT (CE-CT) images of patients treated with sunitinib, a receptor tyrosine kinase inhibitor commonly used for the treatment of renal cell carcinoma, gastrointestinal stromal tumors, and pancreatic neuroendocrine tumors (30). Differentiation between other malignant and benign tumors such as intraductal papillary mucinous neoplasms (IPMN) is also possible through CT feature analysis (31).

The ability to distinguish benign and malignant tumors are important for pre-treatment assessments but for patients already undergoing therapy, identifying genomic phenotypes and their response to therapy is even more important for personalized and effective treatments. If the status of genomic mutations like vascular endothelial growth factor (VEGF), anaplastic lymphoma kinase (ALK), and epidermal growth factor receptor (EGFR) are known before or during treatment, the therapy can be intensified or deintensified to address those mutations specifically for better patient outcome and reduced morbidity. It has been demonstrated such identifications are viable and one study showed that high maximum attenuation values of dynamic CT scans are indicative of VEGF mutations and increased microvessel density before treatment (32). However, some mutations, ALK-rearranged (ALK+) and EGFR, are revealed by comparing pre and early/mid crizotiniband gefitinib, ALK and EGFR inhibitor respectively, treatment image features (33, 34). The changes in three features: central tumor location, absence of a pleural tail, and large a pleural effusion were strongly associated with ALK+ status and are highly discriminatory in patients under the age of 60 and those with operable disease i.e. stage IIIA or lower (33). Renal cell gene mutation analysis is in its preliminary stages but von Hippel-Lindau tumor suppressor (VHL), lysine (K)-specific demethylase 5C (KDM5C), BRCA1-associated protein 1 (BAP1), SET domain containing 2 (SETD2), and polybromo 1 (PBRM1) have all been associated with tumor features like tumor margins, nodular tumor enhancement, and gross appearance of intratumoral vascularity (35).

If mid-treatment adjustments are not viable or likely, accurate patient prognosis becomes even more important. Conventional features such as tumor volume and diameter have been used as prognostic indicators of overall survival (OS) for lung cancer but radiomic features offer more prognostic features in addition to OS. For example, run-length gray-level non-uniformity, Laplace of Gaussian (LoG) run low gray level short run emphasis, and stats median were all statistically significant features for patient survival, while features like wavelet LLH, a wavelet transform sequence involving low, low then high pass filters with down sampling applied to different dimensions of the data, stats range were associated with distant metastasis (DM), and other features were associated with locoregional recurrence (LRR) (20, 36-38). Similarly, in esophageal cancer, tumor texture became more homogeneous after neoadjuvant chemotherapy treatment with a significant decrease in entropy and increase in uniformity features (39).

## 3.2 Positron Emission Tomography

Positron emission tomography (PET) imaging is often used in cancer to survey the metabolic functions of tumors. The significant difference between PET and other imaging modalities is that the radiation originates from the patient through the absorption of radionuclides in the body. One of the most common radiotracer used in PET imaging is fludeoxyglucose (FDG) (40) as a glucose analog. Many cancers, due to their aggressive growth and irregular vasculature, have hypoxic areas that are associated with the reliance on glucose for energy (41). This relationship is further strengthened by the results of a study showing that FDG uptake value derived metagene features were the most prognostic in resected non-small cell lung cancer (42).

In retrospective studies of esophageal cancer patients, high mean standard uptake value ($SUV_{mean}$) and maximum standard uptake value ($SUV_{max}$) were associated with poor disease-free survival (DFS) for patients treated with radiation therapy and lower three-year survival rate for patients that underwent surgical resection of the tumor respectively (43, 44). While SUV values have predictive powers on patient outcome, the correlation of standard TNM staging methods with $SUV_{max}$ and tumor heterogeneity features, entropy and energy, show that radiomic features can compete with current prognostic models and predict non-response, partial-response, and complete-response to therapy with textural features having higher sensitivity than any SUV features (45, 46).

Furthermore, it has been demonstrated with esophageal cancer patients undergoing chemotherapy that temporal changes in second and higher order statistical features had better correlation with pathological response and overall survival than SUV features (46). Similarly, lower coarseness, high contrast and busyness, total lesion glycolysis, and high-intensity emphasis showed better patient outcomes in NSCLC patients (14, 47). Additionally, a regional heterogeneity feature, zone-size nonuniformity (ZSNU), was identified as an independent predictor of progression-free survival (PFS) and disease-specific survival (DSS) in oropharyngeal squamous cell carcinomas (48). Their risk stratification system performed better when ZSNU was combined with other prognostic parameters, total lesion glycolysis and uniformity, a local texture parameter. This suggests that textural features, second and higher order statistical features, describing tumor heterogeneity can provide more prognostic power than just first order SUV features.

### 3.3 Magnetic Resonance Imaging

Magnetic resonance imaging (MRI) can be used for both functional imaging like PET as well as anatomic imaging like CT with the additional benefit of not using ionizing radiation. It has its disadvantages that will be discussed later but one of its advantages over CT is its ability to differentiate tissue types especially in cases of brain cancer where CT scans would have difficulty in identifying tumors within the brain. This advantage of MRIs allow for the use of conventional features like volume in the brain where MRI-derived volumetric features are significantly associated with and are predictive of several cancer-relevant, drug-targetable DNA mutations in glioblastoma as well as the differentiation between mesenchymal (MES) glioblastoma (GBM) and non-MES GBM (49, 50).

This ability to discriminate between tissue and the use of radiomic features allow for advanced identification of tumors and their characteristics. Compared to conventional image analysis, 3D texture analysis approaches allow better discrimination between necrosis and edema versus solid tumor increasing the specificity and sensitivity of brain tumor characterization (51). Tumor differentiation through feature analysis has been observed as early as 1995 in mutual comparison of all tumor types with even specific tumor being successful in certain cases (52). In addition, multigene assay recurrence scores, pathologic response, and semantic features were all significantly associated with radiomic features in brain, breast, and prostate cancers with multi-parametric (MP) MRI feature models excelling in identifying tumors and predicting pathologic response than conventional models (53-57). Tumor contrast enhancement and mass effect also predicted the activation of specific hypoxia and proliferation gene expression programs respectively and showed that the phenotypic expression of these genes have multiple foci for GBMs and are related to significantly shorter survival (58).

Tumor heterogeneity, especially in MRIs continues to be an important factor in determining patient prognosis. Papers have discussed that various MRI derived features and values such as maximum full width at half maximum (mFWHM), entropy, metabolic tumor volumes (MTV), can allow for the identification of tumor heterogeneity in multiple breast and brain cancer types (59-61). The identification of heterogeneity features and others allow for prognostic evaluations to take place. For brain cancer, a subset of ten radiomic peritumoral brain zone (PBZ) MRI features and the histographic pattern of normalized cerebral blood volume were found to be predictive of survival as compared to other features such as enhancing tumours, necrotic regions, known clinical factors, and percent change of skewness or kurtosis (62, 63). For stage IV head-and-neck-squamous-cell-carcinoma (HNSCC) patients with nodal metastases K-trans, a measure of permeability or blood flow, was the strongest predictor of progression-free survival (PFS) and overall survival (OS) (64) while T2w MRI Haralick features were strongly associated with biochemical recurrence following prostate cancer radiotherapy (65). In T2-weighted MRIs of prostate cancer, two textural features, contrast and homogeneity, were compared to the classic apparent diffusion coefficient (ADC) metric and were found to be better predictors of Gleason scores indicative of poorer patient prognosis (66). Additionally, it has been demonstrated in Viswanath et. al. (67) that central gland and peripheral zone prostate tumors have significantly different radiomic signatures alluding to the necessity of using tumor specific features for any classification or prediction problem.

While baseline or pre-treatment prognosis is important, identifying treatment response will quickly become the future of personalized medicine. It's been shown that intratreatment features, like percent change in ADC, change in skewness or kurtosis, large area emphasis, and relative signal intensity AUC, have high potential to predict treatment response in patients with HNSCC, brain cancer, and breast cancer (68-70).

**3.4 Other modalities**

Radiomics is the extraction and association of image features with patient data. With multi-modality methods, the density of information that can be mined increases thus allowing a more thorough analysis of the data presented. PET/CT is one common method where blood flow visualization with DCE-CT was significantly associated with FDG-PET metabolically active volume and uptake heterogeneity for patients with stage 3/4 colorectal tumors (71). Similarly, four features extracted from CT and one from PET were significantly prognostic of distant metastases in NSCLC, the latter of which was also the most prognostic (72). This was supported by a study showing that texture features extracted from fused scans significantly outperformed those from separate scans in terms of lung metastases prediction estimates in PET/MRI images, the best performance being a combination of four texture features extracted from FDG-PET/T1 and FDG-PET/T2FS scans (11). Finally, one feature, co-occurrence of Local Anisotropic Gradient Orientations (CoLIAGe), could be applied to different situations and modalities: differentiation between radiation necrosis vs recurrent tumors in T1w MRI, different molecular sub-types of breast cancer in DCE-MRI, and NSCLC vs benign fungal infection on CT (73).

Another way to increase the information available for analysis is getting temporal information. The big hurdles for this are the increased radiation dose to patients with additional PET or CT images and cost for MRIs. A method to circumvent these obstacles is to use cone beam CT (CBCT). CBCT is a commonly imaging modality used before each fraction of radiation treatment as a tool to check patient positioning. Usually the boney anatomy is used to align the patient and does not require nearly as much radiation as diagnostic images. The low dose and imaging frequency of CBCT make it an ideal candidate to acquire temporal information from despite its poor quality. Fave, Mackin (74) demonstrated that even with the large noise and poor image quality of CBCT images, radiomic features can be extracted from them as long as certain conditions are met. These conditions include: using consistent imaging protocols, relative differences of features, and limiting the patient set to those with less than 1cm of tumor motion. The changes in radiomic features after therapy or delta radioimcs will be discussed later in this review.

For superficial tumors, ultrasound imaging is available and should be considered for radiomic analysis for its quick and non-ionizing properties. It has been demonstrated that an ultrasound signature like gray-level co-occurrence matrix could differentiate radiation induced damage; and an ultrasound method called Nakagami parameters, allows a differentiation between normal and post radiation induced parotid glands and neck fibrosis (24, 25, 75). In addition, sonoelastography, the interrogation of elastic properties of tissue using ultrasound, has been explored as a tool for breast cancer and thyroid malignancy identification (76-78). The radiomics approach was applied to both strain and shear-wave elastography in Zhang et. al. 2017 (78) and Zhang et. al. 2015 (76) respectively and the 2017 paper coined the term sonoelastomics. Both papers demonstrated that radiomic features derived from sonoelastographic images can achieve high AUC, accuracy, sensitivity, and specificity in differentiating breast tumors. The authors go on to suggest that like the recent trend of radiomics, radiomic features in elastography should be applied to predicting normal tissue toxicity and early treatment response.

## 4. Potential applications of Radiomics in Radiation Therapy

The use of radiomics to objectively and quantitatively describe tumors in medical imaging has many potential uses in advancing cancer radiation therapy. It's been demonstrated in this review that radiomic features are valuable and objective sources of information that allow for accurate diagnosis and prediction of patient outcomes. To help further develop the field of radiomics and integrate it to standard of care, many groups have been streamlining the process with automatic segmentation, machine learning, and other novel methods. Simultaneously, other groups have been applying the radiomic process in different modalities and methods such as sonoelastomics, radiomic features extracted from ultrasounds, and delta radiomics, observing the change in radioimc features over time. In short, radiomics would positively impact radiation therapy in three main areas: 1) Reducing disease segmentation time; 2) Reducing disease classification time and providing additional information about the disease classification; and 3) Predicting patient survival curves and normal tissue toxicity.

Automatic segmentation would be useful in clinical oncology in two ways: 1) it would be an objective, reproducible, and standardizable method of finding a region of interest than the subjective human segmentation methods and 2) it would significantly reduce physician disease segmentation time. In CT alone there have been results suggesting that automatic or semi-automatic segmentation methods are as accurate and more stable than manual contouring that yields more reproducible imaging descriptors throughout multiple sites (79-81). Methods like fuzzy locally adaptive Bayesian (FLAB) segmentation are promising automatic segmentation methods (82) that may help in developing computer aided detection (CAD) systems. CAD systems have been successfully tested on lung nodules (83) and are expected to be able to identify many other sites. These radiomic feature based systems complement the qualitative and semi-quantitative radiologists' annotations demonstrating a symbiotic relationship between semantic features and textural features (84). The CAD approach can be applied infields: better assessing cancer risk through more accurate identification of the site and lowering false-positive detection rates (85). One present CAD method involves using texture analysis to analyze the distribution and heterogeneity of SUV and CT values for malignant and benign bone and soft-tissue lesions for improved differential diagnosis on (18)F-FDG PET/CT images (86). These automatic segmentation methods, when trained properly with applicable features, performed admirably with speeds up to 1.54 seconds per slice, about 10-30 seconds per patient, in brain gliomas; significantly faster than manual segmentation methods (87).

These automatic methods require complex algorithms to solve difficult problems and machine learning approaches are becoming more involved as computational power becomes increasingly more powerful and affordable. An example of an ambitious, fully automatic project evaluates the potential of Rad-TRaP framework that comprises three distinct modules: 1: a module for radiomics based detection of PCa lesions on mpMRI via a feature enabled machine learning classifier, 2: a multi-modal deformable co-registration scheme to map tissue, organ, and delineated target volumes from MRI onto CT, 3: generation of a radiomics based dose plan on MRI for brachytherapy and on CT for EBRT using the target delineations transferred from the MRI to the CT (88). This framework could potentially handle a patient from diagnosis to treatment planning and would speed up the pre-treatment process significantly. However, due to the high learning curve of using machine learning in a clinical setting, there are many barriers to the clinicians adopting it quickly. These barriers could be overcome by understanding the primary principles of machine learning (ML) in a clinical setting as outlined in Kang et. al. (89). These primary principles of ML are: 1) data or feature source, dosimetric and non-dosimetric information; 2) supervised or non-supervised feature selection; 3) the ML method used, whether it be random forests, linear regressions, or neural networks; 4) cross-validation methods to generalize the model; 5) testing the generalized model with external data; and 6) comparing with different established models, frameworks to show improvements against the current standard. These principles will allow the clinicians to understand ML at a basic level for discussion and introduction into the clinic. While the clinician's knowledge to apply machine learning methods is important, another limiting factor becomes the library of useful data for the application. A large and detailed library of clinical and non-clinical data would be required for a ML system to work effectively. However, this may not be the case for many studies and could be rectified with the use of either rapid or transfer machine learning methods where both methods use a smaller datasets as a seed to extrapolate and solve the identification, classification, and prediction problems (90) (91).

Some novel methods of extracting radiomic information that could be processed through machine learning include using the surrounding tissue to identify tumors like the heterogeneity of background parenchymal enhancement to identify triple-negative breast cancers (92) or the peri-tumoral fat to correlate pathologically involved axillary nodes (93). Other methods include using radiographic atlases of specific cancer phenotypes to provide insight into niche locations for cancer cells of origin (94). Another possible method is extracting features from novel imaging modalities and radiotracers. Novel modalities like light microscopy can provide a bridge between imaging features and tumor microenvironment (95). Others include the use of novel tracers like 3, 4-dihydroxy-6-(18)F-fluoro-l-phenylalanine ((18)F-FDOPA) to extract features that rely on unique tracer pathways. 18F-FDOPA image analysis has been shown to predict low grade glioma recurrences and look promising as an alternative radiotracer (96, 97).

One of the most important issues in cancer therapy is predicting treatment response in terms of normal tissue toxicity. It is common to build treatment risk models through clinical data, dosimetric evaluation, and semantic image features. Recently, a paper (98) has used clinical, dosimetric, and image based features to build a classification model for normal tissue toxicity. This paper approached the problem of radiation induced shrinkage of parotid glands and the associated long term xerostomia, dryness of the mouth, with machine learning to predict the at-risk patients. This group used a Likelihood-Fuzzy Analysis method that allows the management of heterogeneous variables as well as missing data on a cohort of head and neck cancer patients that underwent radiotherapy. This method allowed the combination of various sources of information: clinical, dosimetric, and radiomic data, to create a classification model predicting the risk of parotid gland shrinkage and 12-month xerostomia. It concluded that while known predictors of normal tissue toxicity were verified, radiomic feature-based models were the best performing.

Predictive models are great for pre-treatment risk analysis but once treatment has ended, it is difficult to characterize the extent of the radiation induced damage due to the discrepancies in physician-based and patient-based assessments. Two papers (25, 99) have suggested two ultrasound based methods: gray-level co-occurrence matric (GLCM) features and Nakagami parameters respectively, to objectively identify the extent of radiation induced fibrosis. GLCM features were used on the basis that normal parotid glands exhibit homogeneous echotexture while radiation induced parotid glands exhibit heterogeneous echotexture. Nakagami parameters are images created from specific statistical distributions, parameters, of the raw ultrasound signals to highlight the underlying tissue structure. Both studies showed significant differences in the ultrasound GLCM features and Nakagami parameters between normal and post-therapy parotid glands representing radiation induced fibrosis.

Most of the current literature has been about predicting clinical endpoints, usually overall survival or normal tissue toxicity, from pre-clinical information. However, recent papers have suggested and explored the concept of delta radiomics. Delta radiomics is a concept coined early 2016 observing that the change in radiomic features during treatment may have information about early chemotherapy treatment response for NSCLC (34, 100). Before it was coined in 2016, other papers have already explored the changes in radiomic features before and after treatment for metastatic renal cell cancer, colorectal liver metastases, and radiation pneumonitis (101-103). Two of those papers, (101, 102), concluded that the changes in two features: entropy and uniformity, representing texture irregularity and homogeneity respectively, were better at treatment response evaluation than the Response Evaluation Criteria In Solid Tumors (RECIST) standard. The other paper, (103), found 12 textural features changed significantly for patients with radiation pneumonitis (RP) grade 2 or above; five of which were significant for grade 3 RP or above. The results suggest that delta radiomics could allow for stratification models for different grades of RP.

The development of delta radiomics in recent years is exemplified by two papers. The first, Aerts, Grossmann (34), was a retrospective pilot study to see if radiomic features could identify gefitinib responders, patients with EGFR mutations, in NSCLC patients. The study compared features from pre-treatment and post three-week treatment to see which features or the change in features was the most predictive of EGFR mutation. From the 183 features extracted from both images, only pre-treatment Laws-Energy 10 and delta volume, maximum diameter, and Gabor Energy features were significantly associated with mutation status. Physical tumor features like volume and maximum diameter have been historically correlated with overall survival, metastasis status, and now mutation status; they may not be an early response feature. However, Laws-Energy 10 and delta Gabor Energy has the potential to add predictive value to the current models of treatment response and may be seen earlier than current indicators. Expanding on this pilot study (104) used delta radiomic features from weekly 4DCT images to interrogate the effectiveness of the delta features. The study concluded that while the selected radiomic features changed significantly on a dose or fraction basis, delta radiomic features did not add significant value to pre-treatment feature models. Some possible cause for the low impact of delta radiomic features are: delta feature selection and model building. The delta features were initially selected from those that had some prognostic value in pre-treatment images and the delta radiomics model was built by adding to other models. This may have culled features that are not prognostic in one image but are prognostic in their changes along treatments and may have reduced the impact of a purely delta radiomics prognostic model. There are many implications which will be discussed later.

Additionally, a group has shown that radiomic features are promising in differentiating patients with and without Meniere's disease (MD), a disorder of the inner ear that can cause vertigo, tinnitus, and hearing loss (105) which could mean that radiomic features could be identified for other diseases and conditions with previously hidden phenotypic patterns.

## 5. Challenges of radiomics

Although there are many applications of radiomics, there are also many challenges to it. Most of these challenges are based on the quality of the image data that may be inherent in the modality itself such as the poor resolution of PET images or the lack of standardization of imaging techniques, parameters, and models that vary from institution to institution. These differences can include: device specifications between brands and models, image acquisition parameters such as image resolution, slice thickness, timing of contrast agent use, etc.; and volume segmentation methods. The effects of different imaging protocols, devices, and parameters are not fully understood and may introduce unknown errors that reduce the validity of the correlations derived from radiomic analysis and many studies (106-108) have called for more research into feature robustness. While these standardization problems could be addressed using robust ML models like neural networks (108) and/or robust features like co-occurance matrix and histogram features (106, 107), the authors believe that a communal effort in creating, testing, and identifying robust radiomic features across machines, models, protocols, and parameters are necessary for the acceleration of radiomic input in the clinical setting.

For all modalities, segmentation is essential to the radiomics process as the systems and algorithms will only look at the regions of interest outlined. However, inter-observer variability depends significantly on the institutions' protocols, observer preference, and requires a significant time commitment for quality segmentation. For example, the common process of using various thresholding can affect segmentation especially in those modalities with low resolution and/or high noise (109-111). The significant variability insists consistency measures, introducing the need for semi- or automatic segmentation processes as mentioned in section 4. When compared to manual segmentation, a semi-automatic segmentation using 3D slicer, image analysis software with interactive segmentation, improved feature quantification reproducibility and robustness (80). Fully automatic segmentation methods are available, like one mentioned in Zhang et. al. (112) that used a random forest classifiers to segment a background and four tumor regions to extract features from them.

In CT, peak tube voltage, tube current, voxel size, image discretization, Hounsfield unit threshold, contrast enhancement, slice thickness, and convolution kernel are among the most basic parameters that affect radiomic features (113-115). These parameters should be standardized for consistent radiomic analysis across institutions and protocols within reason as some parameters such as peak tube voltage are more robust to variability than tube current (115). Along with those basic parameters, there are a few more parameters that can affect radiomic features such as average intensity projection (AIP), free breathing (FB), and end-of-exhale-phase where AIP images contained more DM features than the rest (115, 116). While features can vary, many features like filtered entropy and uniformity are robust to the variation in breathing phase and averaged sets that suggest their strong representation of the tumor (115). Similarly, respiratory gated PET images reduced smearing and improved the quantitation of FDG uptake in lesions (42).

As mentioned before, because of the poor spatial resolution and high noise in PET (109), changes in discretization of image values can greatly affect the textural features (117). This was observed in (118) where grid size had a larger impact on features, like cluster shade and zone percentage, than image iteration number and FWHM. Smoothing does not affect the measurements much (119) but once again, quantization parameters have large effects on the precision of heterogeneity measurements (120). Interestingly, there seems to be a lower limit of how small a volume can be for accurate feature extraction. Tumor volumes less than 45cm$^3$ significantly affected intratumoral uptake heterogeneity comparisons in PET (121).

For multi-modalities like PET/CT or MRI/CT for perfusion and diffusion studies can be difficult to register due to the distortions of the volume of interest (122). An effective registration algorithm is required to minimize registration error (123). A technique using Pearson correlation has shown that instead of using mutual information and correlation ratios as metrics for cost function analysis, a weighted local Pearson coefficient improved significantly (124).

Other errors include study design errors in which type-1 errors, false positives, are too prevalent for accurate association of textural features derived from images to patient survival (125). To minimize type-1 errors, a few methods are available: Benjamini-Hochberg correction (126), bootstrap method correction (127), and if the sample is small enough, an algorithm parameterized by the variance of the spreading distribution can find features robust relative to the spread of the sample (128).

While tumor type, segmentation, modality, and study designs can affect the strength of the correlations derived from radiomic analysis and require careful observation, the quality of the extracted features should be validated to be robust and consistent across those variables. The robustness and consistency of these features can be evaluated by observing the changes due to geometric transformations of the regions of interest and variability of intensities (129). Another method of producing the most significant features is to reduce the number of redundant features through principal components analysis (PCA) along with the Rough set approach to extract the best classification features (130).

Lastly, one of the fundamental principles of the scientific method is reproducibility. However, it is difficult to reproduce results when proprietary or in-house software is used to extract feature. The differences in the software may be minor and can range from the specific names of features or the parameters in which they are pre-processed, and calculated but can have a debilitating effect on the verification of published studies. A recent review showed that there are several software available for each part of the radiomic workflow such as 3DSlicer, MIM, itk-SNAP, LuTA, and Velocity ROI just for image segmentation each with various capabilities and differences (131).

## 6. Discussion

Radiomics, at most, has played an observatory role in oncology so far. The majority of the literature is based on retrospective data that yield specialized features or signatures for specific cancers. While some papers have suggested the translational ability of certain features across cancer types like lung cancer and head and neck cancer (19), there hasn't been a consolidated list of features for cancers as recommendations for future studies or indications for eventual use in clinical practice. Much of this can be attributed to the lack of standardization in extracting radiomic features. A solution may be using a centralized or universally used software platform like the Computational Environment for Radiotherapy Research (CERR) that can import, display, analyze, and create filetypes of treatment plans from different institutions and protocols for easy sharing of information (132). While CERR is a software platform for radiation treatment plans, there have been attempts like the Chang-Gung Image Texture Analysis (CGITA) (133) and MaZda (134), developed for PET and MRI respectively, that can specifically import, display, analyze, and create radiomic features. Most recently, a more general and flexible radiomics software platform, Imaging Biomarker Explorer (IBEX), has been developed at MD Anderson for various applications and aptitudes in radiomics (135).

While the field of radiomics stemmed from the diagnosis of cancers and identification of prognostic factors, there should be a shift towards clinical applications. As therapists, radiologists, and researchers the ultimate goal is to improve the quality of life and patient treatment through the application of research. Much of the published work has laid the groundwork to demonstrate the prognostic power of radiomic features as a non-invasive and repeatable interrogation method for cancers and is easily integrable to the standard of care. As mentioned in recent papers, Aerts (136) and

Fave, Zhang (104), the integration of radiomics into clinical practice depends largely on three factors: adding to the response prediction of current models, earlier treatment response identification, and standardization of radiomics starting from image acquisition to feature extraction. While radiomics is a very hot topic, the lack of standardization from imaging acquisition to the naming of features present a difficult barrier from it going to the next level. However, groups like the Quantitative Imaging Biomarker Alliance from the Radiological Society of North America and Quantitative Imaging Network are working to find and standardize optimal imaging parameters and protocols (136). Additionally, software like Imaging Biomarker Explorer (IBEX) work to create a centralized but modular radiomic software platform to allow consistent data sharing among groups and institutions (135). The ability to share, compare, and verifying the features and conclusions of studies are an important but low impact work that needs to be done.

## 7. Conclusion

Radiomics promises a quantitative solution to the problem of cancer therapy by improving the characterization of tumors through additional datasets and provide further insights into the diagnosis and patient care decision making process. Diseases like glioblastomas and prostate cancer that are very heterogeneous do not yield as much information as homogeneous cancers in biopsies. A non-invasive and repeatable method of interrogating the tumor volume, like medical imaging, for phenotypic information is invaluable for pre-therapy risk analysis. In addition, an advantage of radiomics is that it will not add more radiation burden to patients as it is a field that post-processes images that are part of routine standard of care.

Advances in gene sequencing, phenotypic correlation to clinical outcomes, and machine learning promises a future in which personalized medicine with high sensitivity and specificity. Addressing the limitations of radiomics is the next step in which an emphasis on gathering quality data, extracting the most adequate features, and thorough validation is the hallmark of a good radiomic study. The assessment of radiomic data and models must be driven to be clinically validated, applicable, and universal outside of academic study. This has already been started by the recent TRIPOD (Transparent Reporting of a multivariable prediction model for Individual Prognosis Or Diagnosis) statements in which the clinical usefulness and application of prognostic or diagnostic models of an author's study are reported through a transparent checklist of a study (137). Ultimately, the goal of radiomics is to improve the standard of care by providing more information to healthcare providers in an easily understandable, efficient, and clinically relevant manner.